# The FLARE mission:
# Deep and Wide-field 1-5um Imaging and Spectroscopy for the Early Universe: a proposal for M5 Cosmic Vision call

D. Burgarella*[a], P. Levacher [a], S. Vives [a], K. Dohlen [a], S. Pascal [a], and the FLARE science team
[a]Aix Marseille Université, CNRS, LAM (Laboratoire d'Astrophysique de Marseille) UMR 7326, 13388, Marseille, France

**ABSTRACT**

FLARE (First Light And Reionization Explorer) is a space mission that will be submitted to ESA (M5 call). Its primary goal (~80% of lifetime) is to identify and study the universe before the end of the reionization at z > 6. A secondary objective (~20% of lifetime) is to survey star formation in the Milky Way.

FLARE's strategy optimizes the science return: imaging and spectroscopic integral-field observations will be carried out simultaneously on two parallel focal planes and over very wide instantaneous fields of view.

FLARE will help addressing two of ESA's Cosmic Vision themes: a) « How did the universe originate and what is it made of? » and b) « What are the conditions for planet formation and the emergence of life? » and more specifically, « From gas and dust to stars and planets ».

FLARE will provide to the ESA community a leading position to statistically study the early universe after JWST's deep but pin-hole surveys. Moreover, the instrumental development of wide-field imaging and wide-field integral-field spectroscopy in space will be a major breakthrough after making them available on ground-based telescopes.

## 1. INTRODUCTION

Large ground-based and space-born telescopes have started to scratch the discovery space in the universe, before the end of the reionisation, i.e., the first Gyr of the universe's lifetime. In this era, we will find the very first-light objects, galaxies and black holes. A primordial galaxy, i.e. zero metallicity, that might, contain some pop III stars, that is the first generation of stars created in the early universe at z > 10 [1]. Even more recently, the first solid candidate at z ~ 11, i.e. about 400 Myrs after the big bang, was detected and confirmed by [2].

This latter galaxy is apparently very massive (given the redshift) and rare and wide-field surveys will be necessary to detect them if they happen to be common in the early universe.

On the other hand, the very high redshift of these objects (10 < z < 15) asks for an excellent sensitivity in the near-infrared with a spectral coverage large enough to detect the Lyman break but also fluxes beyond the Lyman break which is at z = 10, the Lyman break is at $\lambda \sim 2$ μm, that is beyond Euclid and WFIRST wavelength range (Fig. 1).

*Denis.Burgarella@lam.fr

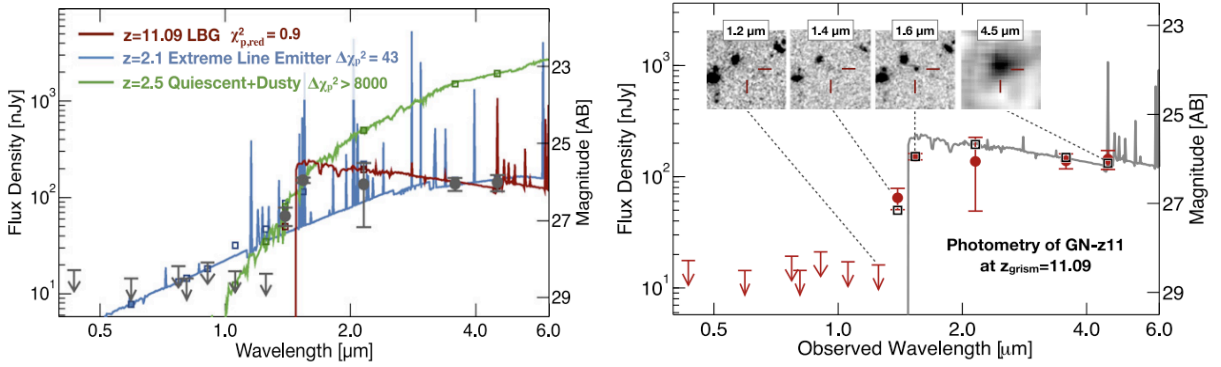

Figure 1. The spectral energy distribution of the galaxy discovered at z = 11.09 [2] presents a discontinuity at λ = 1.6μm. To get a correct and safe estimate of photometric redshifts at 10 < z < 15, it is mandatory to make use of data below and above the Lyman break. The 1 – 5μm wavelength range is therefore a must-do. That is JWST's range but on small fields unlike FLARE. On the other hand, WFIRST, EUCLID and E-ELT will not provide such data.

## 2. BUILDING A CENSUS OF THE OBJECTS AT 10 < Z < 15

The science objectives have been detailed during a workshop http://mission.lam.fr/flare/AgendaMar2016.html.

In the following, we will detail FLARE's science case and why FLARE's capabilities allow to build a census of objects in the early universe up to z ~ 15. FLARE aims at using 3 complementary approaches on the same mission to build a census of the objects at 10 < z < 15. At z = 15, the universe is only 0.3 Gyr old. It is neutral and (probably) sees the formation of the first stars and maybe larger objects.

### 2.1 Detection and identification of a sample of about 100 primordial galaxies at z ~ 15

The expected density of these objects is estimated to be about 1 deg$^{-2}$ at $m_{AB}$ = 28 [3]. To reach this 100-number (Tab. 1), we therefore need to cover at least 100 deg$^2$. JWST is highly unlikely to build surveys much larger than about 1 deg$^2$, i.e., HST-like and will not observe the same type of objects. Moreover, even though JWST will reach magnitudes fainter by about 2 magnitudes, even the E-ELT will not be able to confirm their redshifts spectroscopically because they are too faint. Besides, detecting these primordial galaxies requires any facility to have at least 2 bands at λ > 1.3μm (for z = 10) and λ > 2.0μm (for z = 15). To date, only JWST and FLARE have a wavelength range extending beyond λ = 2.0μm. This sample of galaxies at z ~ 15 will bring invaluable information of the very first phases of galaxy formation but also on larger scale structures. This topic will mainly use the imaging survey, which is the only one to cover an area > 100 deg$^2$. A recent work detected the most remote object in the universe to z = 11.1 [2]. This galaxy is remarkably, and unexpectedly, luminous for a galaxy at such an early time but also very rare. If confirmed, this result implies that the best strategy to detect z > 10 is wide fields, as featured by FLARE.

Table 1. Number density of objects expected at mAB = 28 over 1 sq. deg for No evolution, Empirical and Semi-Analyrical Models (SAM). We conclude that we need about 100 sq. deg. to build a sample of ~100 objects at z ~ 15. From Dark Matter Halo models, the number is much lower, though. But, in this case, even JWST will face a similar issue, or even worse. Extracted from [3].

|  |  | Number Density [objects per 1 deg$^2$] for AB < 28.0 | | | |
|---|---|---|---|---|---|
|  | redshift | No Evolution | Empirical | SAM | DMH |
| 1.0μm-drop | 8-9 | 4,000 | 1,700 | 630 | 850 |
| 1.4μm-drop | 11-12 | 2,400 | 100 | 50 | 4.1 |
| 1.8μm-drop | 14-17 | 1,200 | 0.72 | 1.1 | 0.003 |

## 2.2 Blind spectroscopic survey

Imaging surveys allows detecting galaxies with a strong continuum. However, we know that part of the galaxies, younger and undergoing strong starbursting events are better (only?) detectable via spectroscopic surveys aiming at strong emission lines without any priors coming from broad-band surveys. About 30% of the emission lines objects have no HST counterparts down to I814 > 29.5 (Fig. 2) and the redshift distribution is clearly flatter and reach much higher redshifts [4]. A blind and relatively wide-field integral-field spectroscopic (IFS) survey is the unique way to detect these objects. No other facilities on the sky now or planned will feature such an instrument. FLARE integral-field spectrograph will build a survey via parallel observations and reach magnitudes as deep as the shallow JWST NIRSpec survey but will cover of about 1.5 deg$^2$ for about 500 acrmin$^2$ for NIRSpec (that is 10% only of FLARE to the same limiting flux). Moreover, JWST/NIRSpec will not be blind since a prior photometric detection is needed to define the slits.

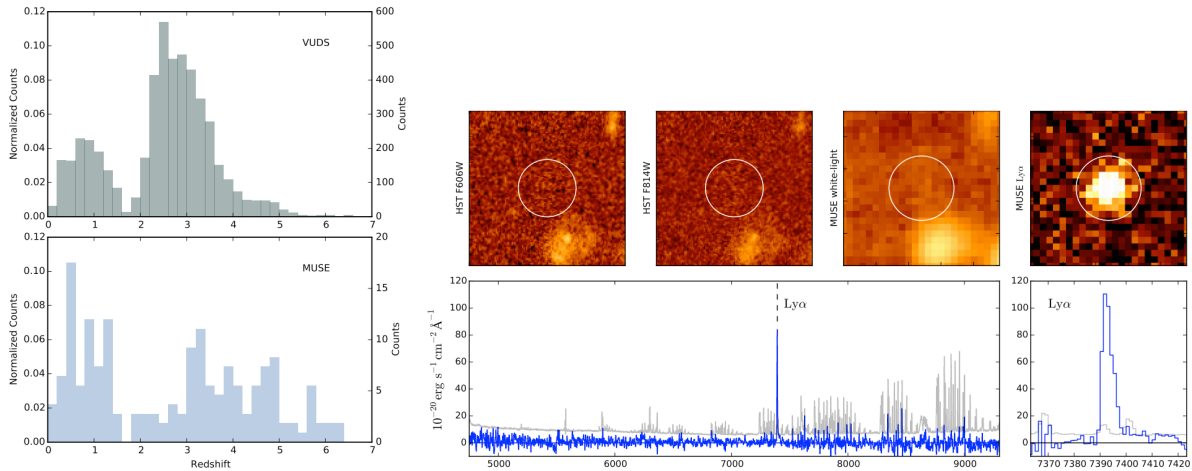

Figure 2. Integral-field spectroscopic observations with MUSE on the VLT [4] in the Hubble Deep Field South allowed to detect as many as 30% of the entire Lyα emitter sample, that have no HST counterparts and thus have I814 > 29.5. Moreover, the MUSE-HDFS (bottom) and VUDS (top) normalized redshift distributions are quite different. This was expected given the very different observational strategy: the VUDS redshift distribution is the result of a photometric redshift selection zphot > 2.3 ± 1σ (with first and second peaks of PDF) combined with continuum selection IAB < 25, while MUSE does not make any pre-selection. With 22% of galaxies at z > 4 in contrast to 6% for the VUDS, MUSE demonstrates a higher efficiency for finding high redshift galaxies. We expect FLARE IFS to produce a large number of sources that will not be detected, even by JWST. ID#553 is a z = 5.08 Lyα emitter without HST counterpart. The HST images in F606W and F814W filters are shown at the top left, the MUSE reconstructed white-light and Lyα narrow band images at the top right. The one arcsec radius red circles show the emission line location. The spectrum is displayed on the bottom figures; including a zoom at the emission line. At the bottom left, the full spectrum (in blue), smoothed with a 4 Å boxcar, and its 3σ error (in grey) are displayed. A zoom of the unsmoothed spectrum, centred around the Lyα emission line, is also shown at the bottom right.

## 2.3 Quasars before the end of the reionisation

Why do we wish to study quasars beyond the end of the reionisation? The four main motivations are:

- To better understand the formation of the first massive black-holes: what are black hole seeds and their early growth?
- To study the environments of the first massive black holes: their metallicity, obscuration, host properties (mass; star formation rate; larger-scale environment), quasar-driven outflows
- Constrain the reionisation of the Universe: contribution to reionisation from quasars and bright sight lines through the reionisation era
- Understand the link between the evolution of galaxies and the evolution of the first black holes as suggested by the similar shape of the redshift evolution of the star formation rate density and the black hole accretion density (Fig. 2).

The density of very high redshift (z > 6 before the end of the reionisation) quasars is low. FLARE's imaging survey is large enough to directly detect 200 of them that we can directly follow up with FLARE integral-field spectrograph. What is unique to FLARE will be the opportunity to detect objects that might be missed by small-field telescopes like JWST but also by wide-field, lower-wavelength and shallower surveys like Euclid and WFIRST via their near-infrared emission and, also, via emission lines.

An important point to stress is that we need to keep in FLARE's observation schedule enough time to directly observe spectroscopically these quasars in the rest-frame ultraviolet-optical frame (Fig. 4) and maybe others that would be detected by ATHENA. They will provide a unique information of the early co-evolution of galaxies and super-massive black holes (Fig. 4) but also, they will allow to study the intergalactic medium on the line of sight.

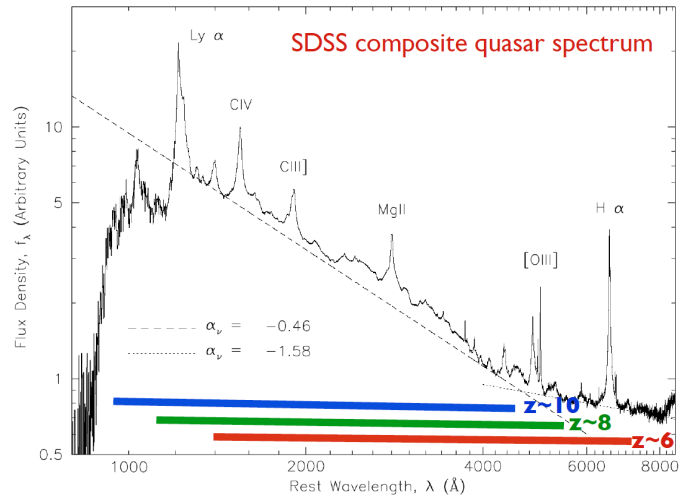

Figure 3. Using [6] we can check that the main ultraviolet-optical lines will be detectable in FLARE's 1-5 μm wavelength range at redshifts z > 6, beyond the end of the reionisation. Of course, this also applies to star-forming galaxies.

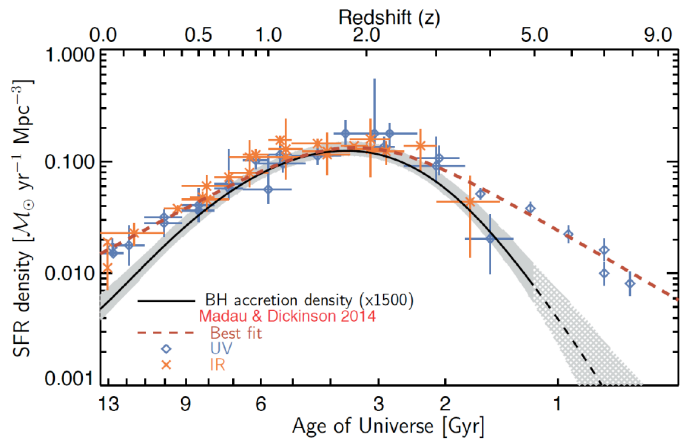

Figure 3. This plot extracted from [5] shows that the Star Formation Rate and the Black Hole Accretion Densities almost present the same evolution in redshift. This suggests a co-evolution of the star formation process along with the black hole one. What happened before the end of the reionisation is of major interest and requires a synergy between several facilities observing quasars at z > 6, ATHENA, FLARE, E-ELT and, of course, JWST.

# 3. STAR FORMATION IN THE MILKY WAY

Herschel and Spitzer allowed to look into the star formation region and to study the interstellar medium (ISM) in the Milky Way. High angular resolution is very important to analyse locally the physics of star formation. But, the dust attenuation in these regions is very high because the young (proto-) stars are still embedded in their molecular, dusty cloud. Both the angular resolution and the wavelength range are needed to « see » deep inside the clouds. Once again, JWST can provide them. But, JSWT's field of view is much too small to cover large areas in the Milky Way

- Unveiling the dark cloud structure and content in the Gould Belt with FLARE: possibility to study dust scattering and extinction on the scale of full regions (Taurus, Rho Oph, Cham, …) in a few days for the small regions, a month for the «Taurus-Perseus-Aurigae» region, less than 6 months total and the possibility to retrieve ≥100,000 stellar spectra with significant extinction in 3 months total.
- 3D Tomography of the Milky Way ISM: FLARE will enable full 3D Tomography of the ISM (Fig. 5) in the entire Galactic Plane down to a spatial resolution of 2 arcmin (at least)
- Census and evolution of Young Clusters in the Galactic Disk : FLARE's superior photometric evolutionary classification of young clusters associated with dense clumps, cross-correlated with other star-formation evolutionary diagnostics, will yield, as a function of Galactic environment, Galaxy-wide measurements of
    - • Cluster formation timescales
    - • Star Formaiton histories
    - • Star formation rates and efficiencies

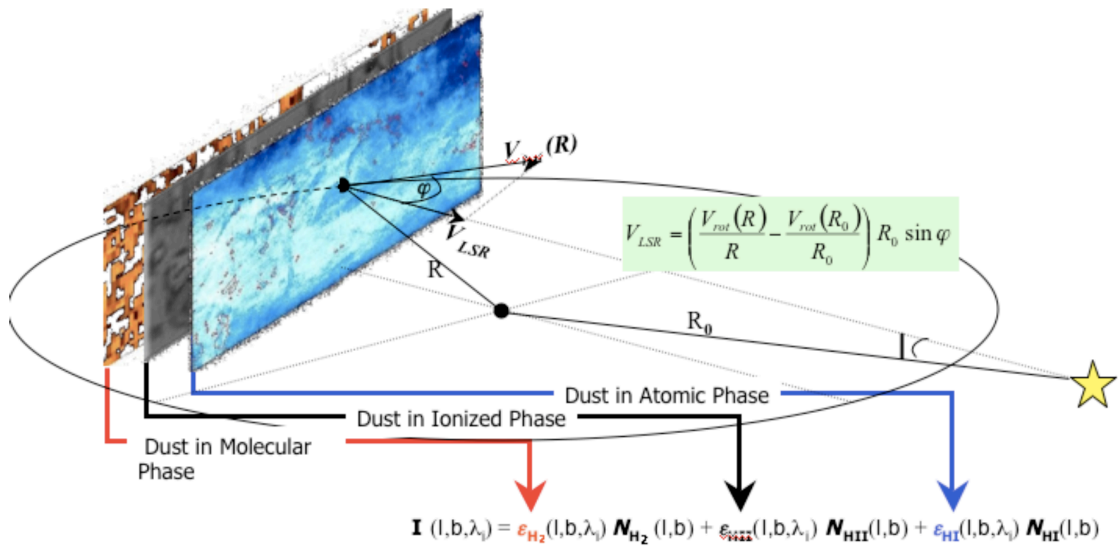

Figure 4. Galactic Inversion to measure the 3-D distribution of dust physical properties in the molecular, atomic and ionized phases of the diffuse ISM [7,8].

# 4. REQUIREMENTS AND OBSERVATION MODES

FLARE will make the best use of the 5-year mission lifetime by observing photometrically and spectroscopically in parallel. This strategy allows to have enough observing time to build, both a wide imaging survey over 100 – 200 sq. deg and a (relatively) wide integral-field spectroscopic survey over 1 – 2 sq. deg. (the size of the present COSMOS survey). To minimize the risks, we will not have any filter mechanisms in FLARE. As illustrated in Fig. 6, the imaging survey will be divided in 6 filters (so far identical to NIRCAM's filters for simplicity by an optimization of the shape needs to be performed). Table 1 summarizing the present requirements assumed for FLARE.

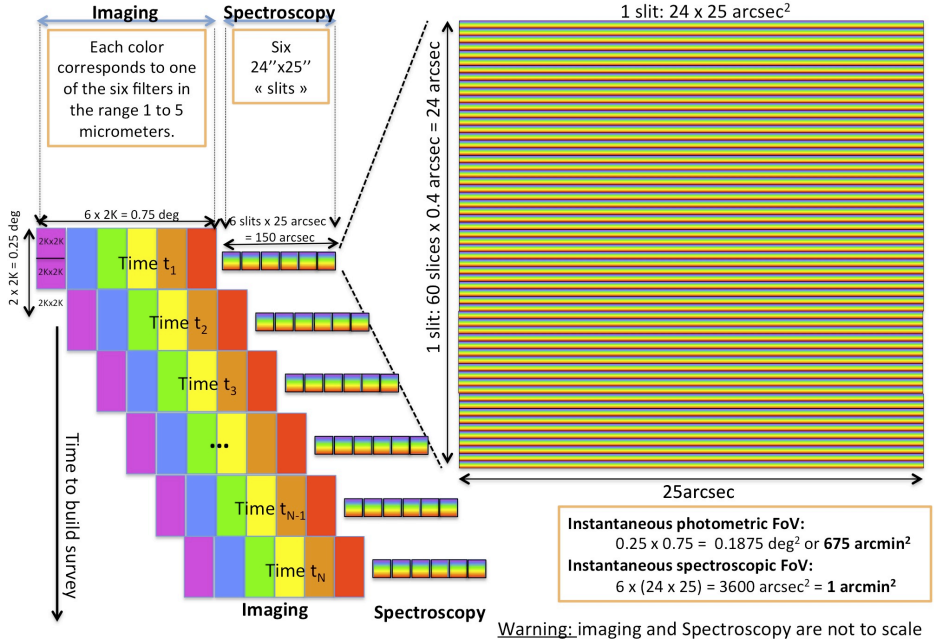

Figure 5. FLARE's strategy combines a wide-field imaging survey along with a wide-field integral-field spectroscopic survey. To make these two surveys efficient, we plan to carry them out in parallel. In addition, targetted observations must be possible to observe rare quasars.

Table 1. Main parameters of the FLARE mission.

| Wavelength range | 1 – 5 μm |
|---|---|
| 6 filters | Identical to JWST NIRCam wide-bands |
| Imaging sensitivity | Better that $m_{AB}$ = 27 - 28 |
| Spectroscopic sensitivity | Better than $f_\lambda = 3 \times 10^{-18}$ erg/cm²/s/Å |
| Primary mirror (M1) | 2.0 m |
| Secondary mirror (M2) | 18.7% of M1: 0.374 m |
| Imaging pixel scale | 0.22 arcsec |
| Transmission telescope | 4 mirrors: $0.97^4$ = 0.885 |
| Encircled energy at first PSF minimum | 0.885 |
| First PSF minimum | 0.441 arcsec @ 3.4um for 2.0m |
| Strehl ratio | 0.90 |
| Spectral resolution imaging R | 4 |
| Length of each slice | 0.4 arcsec |
| Spectral resolution spectroscopy R | 750 |
| Integrated area to compute SNR imaging | [int(0.441 / 0.15)]^2 pixels |
| Collecting area | $\pi [r(M1)^2 - r(M2)^2]$ cm² |
| Pixel scale spectroscopy | 1 slice = 0.4 arcsec |
| Integrated area to compute SNR spectroscopy | 0.4^2 arcsec |
| Transmission spectrograph | 10 mirrors (0.97^10) * grism (0.70) = 0.516 |
| Emissivity for thermal emission | 5% |
| Temperature of telescope | 80K |
| Individual sub-exposure imaging | 500 sec |
| Individual sub-exposure spectroscopy | 500 sec |
| Read-out noise | 15 e- |
| Temperature detector | 50 – 60K |
| Dark | 0.05 – 0.10 e-/sec/pix |
| QE (assumed constant) | 0.75 |
| ADU | 1.53 e-/ADU |
| Total survey time per pixel (imaging and spectroscopy) | ~18000 sec for 2.0m-telescope to reach the depth |

# 5. MISSION AND INSTRUMENT CONCEPT

The proposed FLARE mission consists of a spacecraft module with a large telescope and a payload module including a wide field photometer and an integral field spectrometer. FLARE could be launched by an Ariane 6.2 launcher for injection into an orbit around the L2 Lagrangian point of the Sun Earth system.

As the thermal is a strong driver in the design, and in particular for the detectors, the satellite concept is based on two separate thermal zones: the cold zone integrates the whole optical instrumentation, including the telescope, imager and spectrometers, and the warm zone which is mainly composed by the satellite platform including the electronic boxes used for the command and control of the instrumentation. The thermal dissipation in the cold zone is limited at the best: no mechanisms, only detectors and their front-end electronics are active and will dissipate. These two zones are thermally highly isolated from each other to limit conductive and radiative thermal fluxes between them. During observations, the optical instrumentation zone is fully protected from the Sun by the satellite platform. This configuration authorized a passive cooling of the focal plane assemblies at temperatures close to 40K needed by the detectors for fine observations in the required 1 – 5 µm wavelength range.

The telescope dimensioning and the large number of detectors allow a deep imaging survey with a relatively high angular resolution while preserving a wide field of view. The instrument fields of view are arranged in such a way that the observing strategy will allow to map nearly 200 square degrees in imagery and nearly 2 square degrees in spectroscopy in all spectral bands during the nominal six year life in orbit.

## 5.1 Telescope optical design

The baseline optical concept is based on a 3-mirror Korsch [9] telescope with an effective diameter of 1.8 meters. The combination of the primary and secondary mirrors looks like a Cassegrain configuration forming a real image just behind the primary. This intermediate image is re-imaged by a tertiary with a magnification close to one. This Korsch configuration forms an achromatic image limited by diffraction over a planar annular field of more than 1 square degree.

Mirrors must be in a material able to guaranty a good WFE when cooled down at the low working temperatures. Although the baseline for the primary mirror diameter is 1.8m, this parameter is not currently frozen, and it could vary in the range 1.5 – 2.0 meters. Indeed, it will be optimized in order to achieve the best possible sensitivity, while preserving the overall cost of the mission. Figure 7 illustrates this optical design for a telescope length of 2.8 meters.

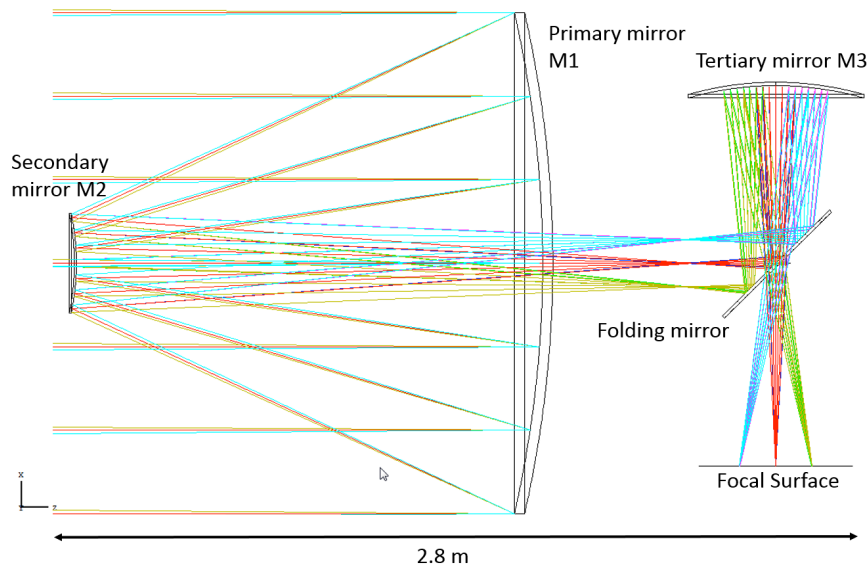

Figure 6. The optical layout of the proposed Korsch design for the telescope of the FLARE mission.

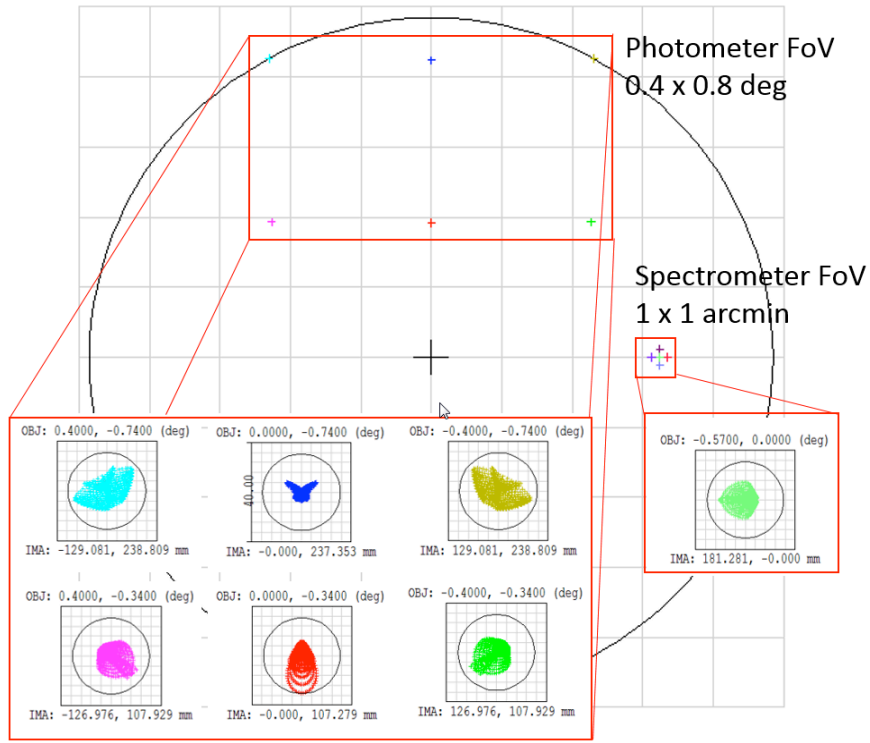

Figure 7. Footprint of the telescope plane showing the image quality (spot diagrams) for the fields of view of the photometer and the spectrometer channels.

Table 2. Optical characteristics of the Korsch-type telescope of FLARE.

|                  |                     |         |
|------------------|---------------------|---------|
| **Primary Mirror**   | Diameter            | 1 800 mm |
|                  | Radius of Curvature | 4040 mm |
|                  | Conic               | -0.98   |
| **Secondary Mirror** | Diameter            | 360 mm  |
|                  | Radius of Curvature | 870 mm  |
|                  | Conic               | -1.80   |
| **Tertiary Mirror**  | Diameter            | 625 mm  |
|                  | Radius of Curvature | 1100 mm |
|                  | Conic               | -0.61   |

## 5.2 Photometer

The photometer consists of 12 (possibly 10) HgCdTe, 2kx2k pixels, infrared detectors located at the focal plane of the telescope mainly for observations in a 15 degree half cone around the zodiacal poles through fixed broadband filters distributed in the 1 - 5 μm wavelength range. Its total field of view covers about 0.16 square degrees.

As shown in Fig. 6, each detector is dedicated to a specific color (with 2 detectors per color).

## 5.3 Integral Field Spectrograph (IFS)

FLARE will also have spectroscopic capability with an integral field spectrograph (IFS) working in the same spectral regions than the imager (i.e. 1.25 – 5 µm) with an instantaneous field of view of about 1 arcmin$^2$. The IFS consists of 2 identical arms each made of three main modules: the Fore-Optics unit, which formats the 2D entrance FoV; a set of 3 Image Slicer Units feeding 3 Spectrograph units. Each spectrograph has two spectral channels in order to accommodate the two octaves of the wavelength range. The total spectral range of the IFS prevents for inserting lenses or dioptric elements within the overall optical layout. An all-reflective design takes advantage of its facility to adapt to a cryogenic environment and presents a higher throughput.

The Fore-Optics (FO) unit re-images the input F/10 telescope field onto the slicing mirrors and introduces an anamorphic magnification of the field (F/130 in the spectral direction; F/65 in the spatial direction). The anamorphous allows squeezing the spatial direction of the FoV to limit the length of the slices. The Fore-Optics unit is composed of four re-imaging mirrors (FM1 to FM4) plus one folding mirror as shown in Fig. 9. The two mirrors FM1 and FM3 are toroidal and perform the anamorphous of the beam by a factor 2 in one direction. The secondary is elliptical while the forth mirror is spherical. Despite the apparent complexity of the use of four mirrors this keeps the manufacturability and allows reaching high surface quality of each component.

Each Fore-Optics accepts a larger FoV of 1x2 arcmin (instead of 1x1 arcmin required) in order to accommodate the 6 Image Slicer Units.

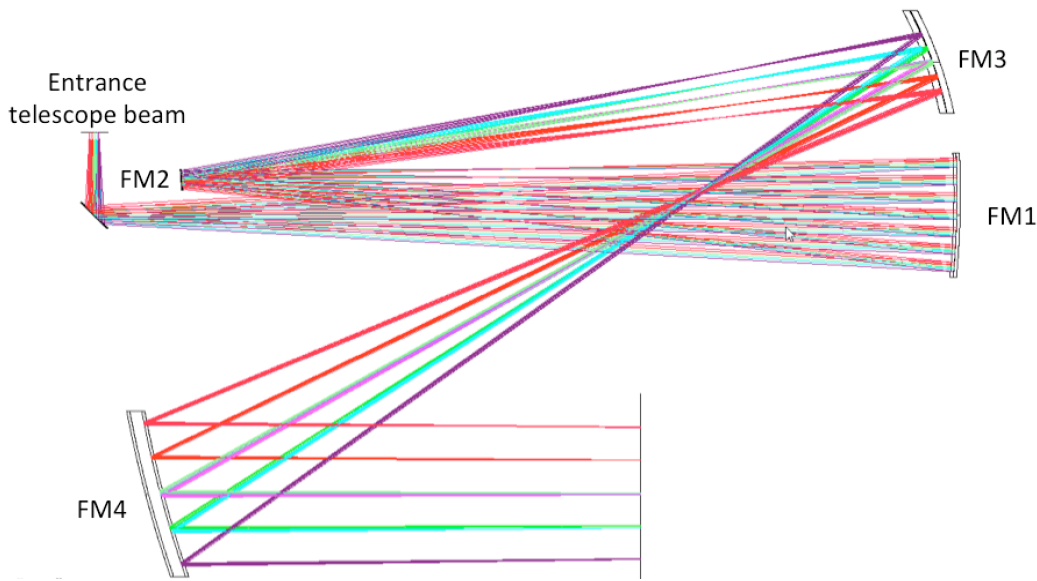

Figure 8. Optical layout of the Fore-Optics Unit.

Each Image Slicer Unit re-arranges the 2-D FoV at the input focal planes of one spectrograph to form two entrance slits for the Spectrograph.

One Image Slicer Unit consists of three optical assemblies: the two slicing mirror arrays at the output image planes of the Fore-Optics, an array of pupil mirrors and an array of slit mirrors. Fig. 10 shows 6 channels (i.e. 6 slices) of one Image Slicer Unit.

The slicing mirror array consists of two stacks of 28 (56 in total) concave spherical mirrors (0.5 mm wide and 12 mm long), which slice the anamorphic field and produce an array of individual images of the pupil on the pupil mirrors. The proposed method for the manufacturing of the slicing mirrors is by glass polishing. Slices are made in Zerodur and assembled by optical contacting. The current design of the slicer (stack of 28 slices) is compatible with an innovative method [10] enabling the manufacture of one or more stacks of slices by a single standard polishing process thus reducing both the time and cost of production.

The pupil mirror array consists of four staggered lines of 7 rectangular mirrors. Each pupil mirror re-images its own slice of the anamorphic field on to a dedicated slit mirror located at the input focal plane (slit plane) of the Spectrograph. The slit mirror array consists of a single line of 28 rectangular mirrors each with 2 mm width. Each slit mirror re-images the telescope pupil on its pupil mirror to the entrance pupil of the Spectrograph Unit. The surface of each slit mirror is spherical and concave.

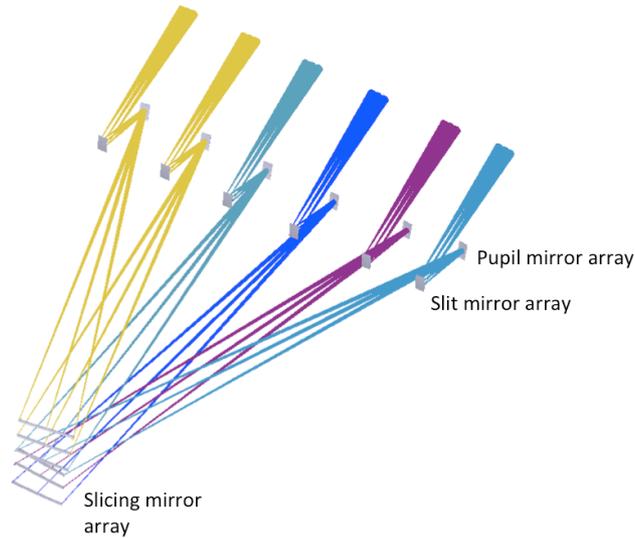

Figure 9. Optical layout of the Image Slicer Unit (only 6 slices are shown).

Each spectrograph unit is based on an Offner configuration with a magnification of 0.5. At the entrance of the spectrograph, a dichroic is located to split the octaves of the wavelength range. The secondary mirror acts as the grating and produces spectrum with a spectral resolution higher than 500 on an HgCdTe, 2kx2k pixels, infrared detector, identical to the one of the photometer. Each spectrograph accepts two entrance slits, which are dispersed and imaged on a common detector.

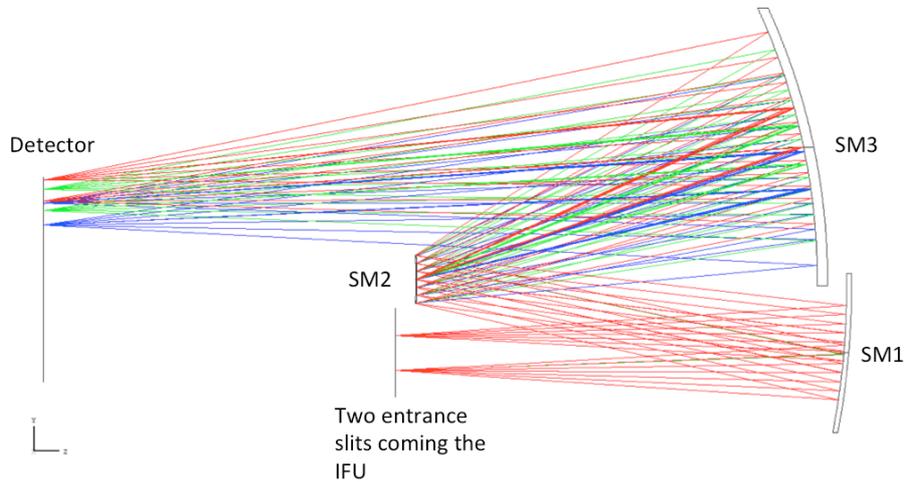

Figure 10. Optical layout of the Spectrograph Unit. The dichroic is not shown.

# 6. CONCLUSION

FLARE can open up a new domain of exploration for the early universe during the first giga-year or the universe by proving simultaneously deep photometric and integral-field spectroscopic data over wide fields (100 - 200 sq. deg in photometry and 1 - 2 sq. deg. in integral-field spectroscopy) in the near-infrared range up to 5μm. No other projects in operation or planned present the same capabilities. It offers a new opportunity that is complementary to the other missions a) wide-field, $\lambda < 2\mu m$ and b) small fields, $\lambda > 2\mu m$.

FLARE's strategy will be optimized to build a census of the objects residing in the very early universe, namely: continuum-bright objects like Lyman break galaxies using the wide-field imaging survey, emission-line objects like Lyman alpha emitters using the 1 - 2 sq. deg integral-field spectroscopic survey, and, a sample of at least 200 quasars.

Beside this early universe facet that will be dominant with about 70-80% of the mission lifetime, we plan to keep ~20% to observe star-forming regions in the Milky Way, and, finally some parts of the mission could be dedicated to targeted observations in synergy with ATHENA, SKA or the E-ELT.

# REFERENCES

[1] Sobral et al. (2015, ApJ 808, 139)
[2] Oesch et al. (2016, ApJ 819, 129)
[3] Yamada et al., Proceedings of the SPIE, Volume 7731, p.77311Q (2010)
[4] Bacon et al., A&A 575, A75 (2015)
[5] Aird et al., ApJ 815, 66 (2015)
[6] Vanden Berk et al., AJ 122, 549 (2001)
[7] Paladini et al. 2007,
[8] Traficante et al. 2014
[9] D. Korsch, "Anastigmatic three-mirror telescope" Applied Optics 16, pp. 2074–2077, 1977.
[10] S. Vives, et al., "New technological developments in Integral Field Spectroscopy", Proc. SPIE 7018, p.70182N (2008)